\documentclass[aps,pra,showpacs,twocolumn,amsmath,amssymb,superscriptaddress]{revtex4-1}

\usepackage[english]{babel}
\usepackage{latexsym}
\usepackage{graphics,psfrag}
\usepackage{epstopdf}
\usepackage{epsfig}
\usepackage{color}
\usepackage{hyperref}
\usepackage[normalem]{ulem}
\hypersetup{
  colorlinks   = true,  
  urlcolor     = blue,  
  linkcolor    = blue,  
  citecolor    = red    
}

\def\be{\begin{equation}}
\def\ee{\end{equation}}
\def\bea{\begin{eqnarray}}
\def\eea{\end{eqnarray}}
\def\bi{\begin{itemize}}
\def\ei{\end{itemize}}
\def\bin{\begin{enumerate}}
\def\ein{\end{enumerate}}

\begin{document}
\title{Rydberg dressing of a one-dimensional Bose-Einstein condensate}

\author{Marcin P\l{}odzie\'n}
\affiliation{Eindhoven University of Technology, PO Box 513, 5600 MB Eindhoven, The Netherlands}
\author{Graham Lochead}
\affiliation{Van der Waals-Zeeman Institute, University of Amsterdam, Science Park 904, 1098 XH Amsterdam, The Netherlands}
\author{Julius de Hond}
\affiliation{Van der Waals-Zeeman Institute, University of Amsterdam, Science Park 904, 1098 XH Amsterdam, The Netherlands}
\author{N.~J. van Druten}
\affiliation{Van der Waals-Zeeman Institute, University of Amsterdam, Science Park 904, 1098 XH Amsterdam, The Netherlands}
\author{Servaas Kokkelmans}
\affiliation{Eindhoven University of Technology, PO Box 513, 5600 MB Eindhoven, The Netherlands}

\date{\today}

\begin{abstract}
We study the influence of Rydberg dressed interactions in a one-dimensional (1D) Bose-Einstein Condensate (BEC). We show that a 1D geometry offers several advantages over 3D for observing BEC Rydberg dressing. The effects of dressing are studied by investigating collective BEC dynamics after a rapid switch-off of the Rydberg dressing interaction. The results can be interpreted as an effective modification of the $s$-wave scattering length. We include this modification in an analytical model for the 1D BEC, and compare it to numerical calculations of Rydberg dressing under realistic experimental conditions.
\end{abstract}

\pacs{03.75.Hh, 32.80.Ee, 32.80.Qk}
\maketitle

Ultracold quantum gas experiments allow for an extremely precise control of interatomic interactions. Strong interactions at the atomic level enable in principle the creation of strongly correlated many-body systems, where the tunability gives them an important advantage over their solid state equivalents. The short-range interactions between ground-state atoms can be controlled by Feshbach resonances 
, which resulted, e.g., in the demonstration of the BCS-BEC type superfluid crossover 
However, while the interactions can be made very strong by going to the unitarity regime, it is still under debate whether this quantum gas can be considered as a strongly correlated system 
. Strong correlations will be evident when the interaction is both strong and long range, i.e., the range of the interaction exceeds the average interparticle separation. Rydberg atoms 
take a central role in the broad spectrum of systems that can be categorized from short range to long range, since Rydberg atoms can be classified as intermediate range. Their mutual interaction is generally of van der Waals nature, which is neither long range or short range: the van der Waals coefficient $C_6 \sim n^{11}$ scales rapidly with the principal quantum number $n$, allowing for a range larger than the interparticle separation.

Rydberg atoms in the context of ultracold atomic gases \cite{lukin2001,Heidemann2007,Urban2009,gaetan2009,weimer2008,pohl2010,bijnen2011,Schachenmayer2010,pritchard2010,Henkel2010,Johnson2010,Honer2010} open up a whole new direction of strongly correlated many body physics with a focus on quantum computation and quantum simulations \cite{jaksch2000,Weimer2010,Pohl2011,Keating2013,Pohl2014,Zoller2015,Pohl2015}. Most of the applications have their origin in the ability to manipulate these Rydberg atoms coherently on timescales below their radiative lifetime. When the atoms are cold enough, they remain spatially frozen at those timescales. Off-resonant coupling to Rydberg states, referred to as Rydberg dressing \cite{Henkel2010,Johnson2010,Honer2010}, allows experimentalists to achieve timescales longer than those required to stay in the frozen gas limit. These timescales enable BEC dynamics with long-range interactions, which is predicted to give rise to novel exotic many-body physics such as supersolidity \cite{Pupilo2010,Boninsegni2012,Henkel2012,Grusdt2013,Mason2012,Kunimi2012,Macri2013}.

Rydberg dressing of individual atoms trapped in optical tweezers \cite{Jau2016} and in optical lattices \cite{Zeiher2016} has been observed experimentally. Furthermore, modification of electromagnetically induced transparency (EIT) via resonant Rydberg dressing has also been observed \cite{Gaul2015}.  However, observation of Rydberg dressing of a regular BEC has proven elusive so far. Rydberg dressing of a BEC in 3D has been theoretically studied \cite{Honer2010,Pohl2011a,Balewski2014,Lesanovsky2012,Wuster2015} where at relatively low density the influence of dressing could be interpreted as a modification of the $s$-wave scattering length.

In Ref.~\cite{Balewski2014}, the authors conclude that experimental observation of a dressed BEC in 3D is very difficult due to a strong reduction in the amount of Rydberg atoms in the BEC caused by the Rydberg blockade mechanism. The characteristic density at which the blockade becomes dominant is very low, compared to typical BEC conditions in 3D, and even at densities below this characteristic value the dressing interaction is already strongly suppressed. Moreover, the interpretation as an effective modification of the scattering length is not obvious under these conditions. 

Here, we consider a one-dimensional geometry, and derive results for Rydberg dressing of 1D BECs.
We show that these lead to much more favorable conditions for the observation of BEC dressing and that a sudden switch-off of the dressing interaction (via a switch-off of the dressing laser) results in collective BEC dynamics, which allows an interpretation in terms of an effective change of the $s$-wave scattering length. We show that the BEC breathing mode is a tenable experimental signature to observe BEC Rydberg dressing, under realistic conditions, e.g.\  those typical for a 1D BEC trapped on an atom chip \cite{Schumm2005,Esteve2006,Amerongen2008}.  

\begin{figure}
\centering
\includegraphics*[scale=0.30]{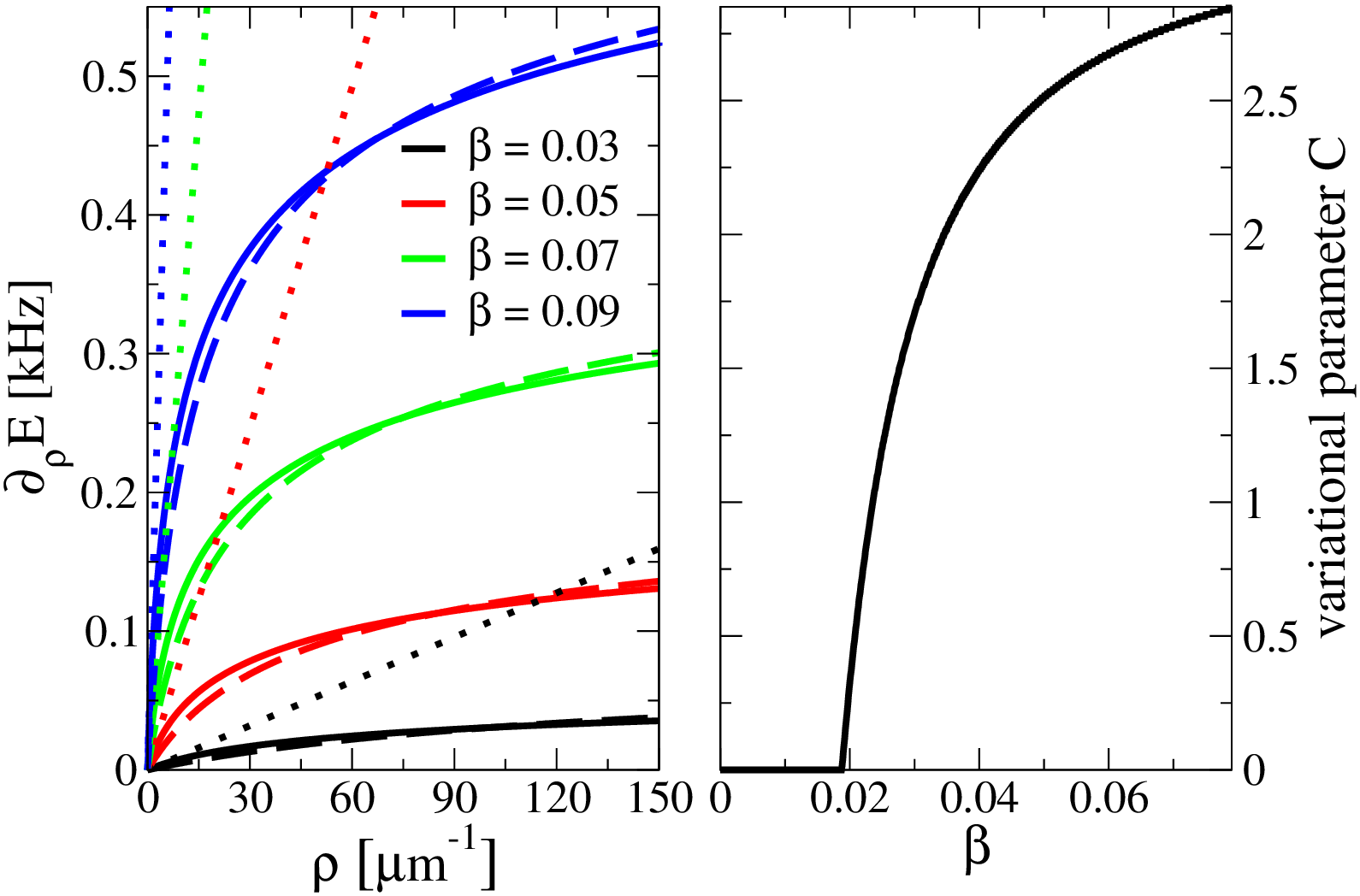}
 \caption{Left panel: Characteristic behavior of the Rydberg-dressed interaction in 1D, as a function of BEC linear density, for realistic experimental conditions, using the $35S$ Rydberg state of $^{87}$Rb. The energy functional $\partial_\rho E[\rho]$ is plotted for different dressing Rabi frequencies, as  parameterized by $\beta=\varOmega/2|\varDelta|$ for fixed detuning $\varDelta/2\pi = - 100$\,kHz. Corresponding critical densities $\rho_B$ are (curves from bottom to top) $177\,\mu$m$^{-1}$, $64\,\mu$m$^{-1}$, $33\,\mu$m$^{-1}$ and $20\,\mu$m$^{-1}$.
 Dashed lines correspond to  the  variational ansatz $F$, Eq.~(\ref{eq:F_ansatz}), dotted lines correspond 
 to a linear increase with density as in the two-body limit. 
Right panel: variational parameter $C$ for the energy functional ansatz $F$. Vanishing $C$ for $\beta<0.018$ corresponds to the two-body limit, Eq.~(\ref{g_two_body_limit}), while $\beta>0.018$ corresponds to  the saturation effect Eq.~(\ref{omega_analytical}).}
\label{energy_dressing}
\end{figure}

The advantages of the 1D geometry are twofold. First, it becomes much easier to work at densities {\em below} the characteristic blockade density. Second, the suppression of the Rydberg-dressed interactions by the blockade mechanism that is present even at low densities in 3D,  is much less severe in 1D. Both of these are evident in Fig.~\ref{energy_dressing}, which shows the behavior of the relevant energy functional for realistic experimental parameters (see below for details).

Before discussing the differences between 1D and 3D for Rydberg dressing of  a BEC, it is useful to first quantitatively introduce the Rydberg-dressed two-body interaction \cite{Henkel2010,Johnson2010,Honer2010}.
 
For the purpose of this paper we limit ourselves to isotropic  Rydberg dressing, by (two-photon) off-resonant coupling the electronic ground state of the atom to a Rydberg $S$ state with  principal quantum number $n$. The key parameters are now the dressing parameter $\beta= \varOmega/2|\varDelta|$ and blockade radius 
$R_B = \left( C_6/2\hbar\sqrt{\varDelta^2+\varOmega^2} \right)^{1/6} \approx \left( C_6/2\hbar|\varDelta| \right)^{1/6}$, where 
$\varOmega$ is the two-photon Rabi frequency, $\varDelta$ is the total (red) detuning and $C_6>0$ is the van der Waals coefficient. The resulting interaction potential between dressed atoms has the form $W(r) = \beta^4\frac{C_6}{R_B^6+r^6}$ \cite{Honer2010}. In practice, the blockade radius $R_B$ is in the $\mu$m range.

While the Rydberg-dressed interaction can be strong because of the high value of $C_6$, for larger densities (typical for a BEC) it saturates around a characteristic density $\rho_B=1/\beta^2V_B$, where there is one excited Rydberg atom per Rydberg blockade volume $V_B$. This leads to an overall offset of the chemical potential of the BEC, but only to a small modification of the shape compared to no dressing.  Only for low relative density $\rho<\rho_B$ is there a significant alteration of the BEC shape due to dressing. In 3D, as considered in Ref.~\cite{Balewski2014}, one has $V_B=\frac{4}{3}\pi R_B^3$ and for typical parameters $\rho_B$ is lower than that of typical densities of 3D BECs. In contrast, in the 1D case we consider here, one has a 1D blockade volume $V_B=2R_B$ leading to a different scaling of $\rho_B$.  We find that for practical experimental conditions the entire 1D BEC can have a density below $\rho_B$.

We now give a quantitative description of the effects of the Rydberg dressing on an (effectively) 1D BEC. 
We assume a cigar-shaped BEC, in a trap with a radial trapping frequency $\omega_\perp$ much larger than the axial trapping frequency $\omega_0$ and a chemical potential $\mu<\hbar\omega_\perp$. 
We start from the generalized Gross-Pitaevskii (GP) equation
$i\hbar\frac{\partial\psi(x)}{\partial t}~=~\left[ -\frac{\hbar^2\partial_x^2}{2m} + g_0 N|\psi(x)|^2 + \frac{m\omega_0^2x^2}{2} + V_{MF} \right]\psi(x)$,
where $N$ is the particle number, and  $g_0=2a_s\hbar\omega_\perp$ is the one-dimensional mean-field coupling parameter. 
In the mean-field regime, and assuming the radial size of the BEC is much smaller than $R_B$ (and at sufficiently low densities where collective effects can be neglected)
$V_{MF} =  N  \int W(x-x')|\psi(x')|^2dx'$,
which is treated in the same way as 
recently shown in Rydberg physics \cite{Henkel2010,Pohl2011}. 
If we assume that the length of the 1D BEC is much larger than the Rydberg blockade radius $R_B$, so that the density is constant over this radius, we may approximate the $V_{MF}$ term as  $V_{MF} \approx N |\psi(x)|^2 \int W(x') dx'$.
Contrary to 
the bare van der Waals interaction, the combination of short-range saturation and long-range $1/r^6$ tail of the two-body dressed interaction assures that the above integral is finite, and therefore gives a correction to the 1D mean-field coupling $g_0$ of
 \begin{equation}\label{g_two_body_limit}
g_{\rm eff} = g_0 + \frac{2}{3}\pi\frac{C_6}{R_B^5} \beta^4 = g_0 + \frac{\pi}{12}\frac{\hbar\varOmega^4}{|\varDelta|^3}R_B,
 \end{equation}
and corresponds to a linear increase of $\partial_\rho E$ with density in Fig.~\ref{energy_dressing}. 
Note that when these conditions are applied to a 3D BEC, we similarly find 
$g_{\rm eff}^{3D} = g_0^{3D} + \frac{2}{3}\pi^2\frac{C_6}{R_B^3}\beta^4 = g_0^{3D} + \frac{\pi^2}{12}\frac{\hbar\varOmega^4}{|\varDelta|^3}R_B^3$, 
which has already been derived in Refs.~\cite{Honer2010,Pohl2011a,Balewski2014}.

Next, to allow for higher linear densities where collective effects can play a role, we adapt the treatment of Ref.~\cite{Honer2010} to 1D, again using the assumption that the radial size of the BEC is much smaller than $R_B$. In this way, we obtain an  energy density for the internal degrees of freedom $\epsilon_\mathrm{var}$ at linear density $\rho$ 
\begin{equation}\label{eq:VarEnergy}
        \epsilon_\mathrm{var}(\theta,\xi) = \frac{\hbar \Delta \rho}{2} \cos 2\theta - \frac{\hbar \Omega \rho}{2} \sin 2\theta + \frac{\lambda}{2} \sin^4\theta \frac{C_6 \rho^2}{\xi^5},
\end{equation}
having as variational parameters the Rydberg mixing angle $\theta$  and the correlation length $\xi$. 
Note that in 1D the last term scales as $\xi^{-5}$, compared to $\xi^{-3}$ in 3D. The correlation length $\xi$ is constrained to the blockade radius $R_B$ for low densities, and to the average distance between Rydberg atoms, $1/f\rho$, 
for higher densities, $\xi\approx\min(R_B,1/f\rho)$, with $f=\sin^2\theta$ the Rydberg excitation fraction. In this approach, minimizing the energy density with respect to $\theta$ yields the energy functional $E[\rho]$, and the derivative $\partial_\rho E[\rho]$ (see solid lines in Fig.~\ref{energy_dressing}) is inserted as  $V_{MF}$ into the generalized GP equation. In the low-density limit of Eq.~(\ref{eq:VarEnergy}), one has $\xi=R_B$ and $\theta$ approaches $\beta$; comparing the result to the Rydberg part of the two-body limit Eq.~(\ref{g_two_body_limit}) allows us to determine the value  $\lambda=2\pi/3$ in Eq.~(\ref{eq:VarEnergy}).

As a specific example, we consider dressing the $^{87}$Rb ground state with the  $n=35S$ Rydberg state (which has $C_6/\hbar= 2\pi\times 0.1891$ GHz~$\mu$m$^6$  \cite{Singer2005}) and a detuning $\varDelta=-2\pi\times 100$~kHz, leading to a blockade radius $R_B=3.13$~$\mu$m. (Note that for $n<50$  Rydberg dressing will not be
affected by molecular dimer excitations, as the splitting between the
dimer and atomic excitation line is always greater than the typical
linewidth of $<1$~ MHz~\cite{gaj2014}). The resulting energy functional is shown in Fig.~\ref{energy_dressing}, for various Rabi frequencies parameterized by $\beta$, and as a function of linear density $\rho$. For the parameters and densities in Fig.~\ref{energy_dressing} we find $R_B<1/f\rho$, and thus we may take $\xi=R_B$ throughout.  It is worth emphasizing here that the saturation in the energy functional as the density increases is much less severe in 1D when compared to 3D, and that the densities of 1D BECs in Fig.~\ref{energy_dressing} are in the range that is typical for those achieved on atom chips \cite{Schumm2005,Esteve2006,Amerongen2008}. 

A useful approximate description can be obtained as follows: the energy functional $E[\rho]$ is qualitatively described by the anzatz
\begin{equation}\label{eq:F_ansatz}
 F[\rho] = \epsilon_0 - \epsilon_0\left(1+\frac{2}{3}\pi C\frac{\rho}{\rho_B}\right)^{-1/C},
\end{equation}
which fulfills the two limiting cases: 
a) it saturates for $\rho \gg \rho_B $ on a constant value $\epsilon_0 = \hbar|\varDelta|\beta^2$ \cite{Honer2010}; b) it is linear with density for small densities, $\displaystyle{F \approx   \pi/12\hbar\varOmega^4/|\varDelta|^3R_B\rho}$, see Eq.~(\ref{g_two_body_limit}). The free parameter $C$ is numerically obtained by minimizing the  root mean  square of the difference between $F$ and $\partial_\rho E[\rho]$, these results are also shown in Fig.~\ref{energy_dressing} as dashed lines.  
The linearization of Eq.~(\ref{eq:F_ansatz}) around the trap center allows us to extract the effective mean field coupling constant $g_{\rm eff}$. As the Rydberg dressing will only have a small effect on the Thomas-Fermi profile, as will be shown in the following paragraphs, we are able to express the dressing effect as an effective change of the trapping frequency. This is done by equating all the different potential energy terms in the GP equation:
\begin{eqnarray}
g_0 \rho + m\omega^2_0 x^2/2 +F[\rho] &\approx& g_0 \rho + m\omega^2_{\rm eff} x^2/2 \nonumber \\
&=& g_{\rm eff} \rho +m\omega^2_0 x^2/2,
\end{eqnarray}
with $\rho$ the undressed Thomas-Fermi profile. After linearization of $F$ around the trap center, we find the effective mean field coupling constant including the saturation effects of dressing to be
\begin{equation}\label{omega_analytical}
g_{\rm eff} = g_0 + \frac{\pi}{12}\frac{\hbar\Omega^4}{|\varDelta|^3}R_B\left(1+\frac{\pi}{3} C\frac{\rho_{0}}{\rho_B} \right)^{-1/C-1},
\end{equation}
where  $\rho_{0}$ is the BEC peak density and $C$ is a variational parameter for Eq.~(\ref{eq:F_ansatz}). One can observe that the two-body limit is obtained for vanishing densities, as well as for $\beta < 0.018$, see Fig.~\ref{energy_dressing}.

In order to further quantify the effects of Rydberg dressing on a 1D BEC, we now consider a more specific scenario.
We assume a BEC of $N = 200, 1000, 2000$ $^{87}$Rb atoms with an $s$-wave scattering length $a_s = 99a_{0}$ (where $a_0$ is the Bohr radius) confined in an cylindrical harmonic trap with radial frequency $\omega_\perp = 2\pi\times 3000$~Hz and axial frequency $\omega_0 = 2\pi \times 30$ Hz. These parameters correspond to a mean-field coupling strength $g_0 = 2.08 \times 10^{-38}$~Jm $ = 0.53 E_0 l_0$ with $E_0 = \hbar\omega_0$, $l_0 = \sqrt{\hbar/m\omega_0}=1.97~\mu$m, and $t_0 = \omega_0^{-1} = 5.3$~ms as a energy, length and time unit respectively. In the absence of dressing, the corresponding BEC half-length along the axial direction and peak density are $10.6\,\mu$m and $28\,\mu$m$^{-1}$ for $N =200$, $18.2\,\mu$m and $81\,\mu$m$^{-1}$ for $N = 1000$, $23\,\mu$m and $128\,\mu$m$^{-1}$ for $N=2000$. 
 
The  effects of the Rydberg dressing can now be characterized as follows. For different particle number $N$ and dressing parameter $\beta$ (keeping $\varDelta=-2\pi\times 100$~kHz fixed) we first calculate the ground state $\psi_0$ of the generalized GP equation, using $\partial_\rho E$ obtained from Eq.~(\ref{eq:VarEnergy}).
Next, we calculate  the time evolution of this initial wave packet upon a sudden switch-off of the Rydberg dressing and numerically calculate the relative change of the BEC size defined as $\kappa(t) = R(t)/R_0$, where $R_0$ is the initial dressed BEC half-length in the axial direction and $R(t)$ is  the  half-length in  the  axial direction obtained from fitting parabola to the  undressed cloud at time $t$.

\begin{figure}
\centering
\includegraphics*[scale=0.30]{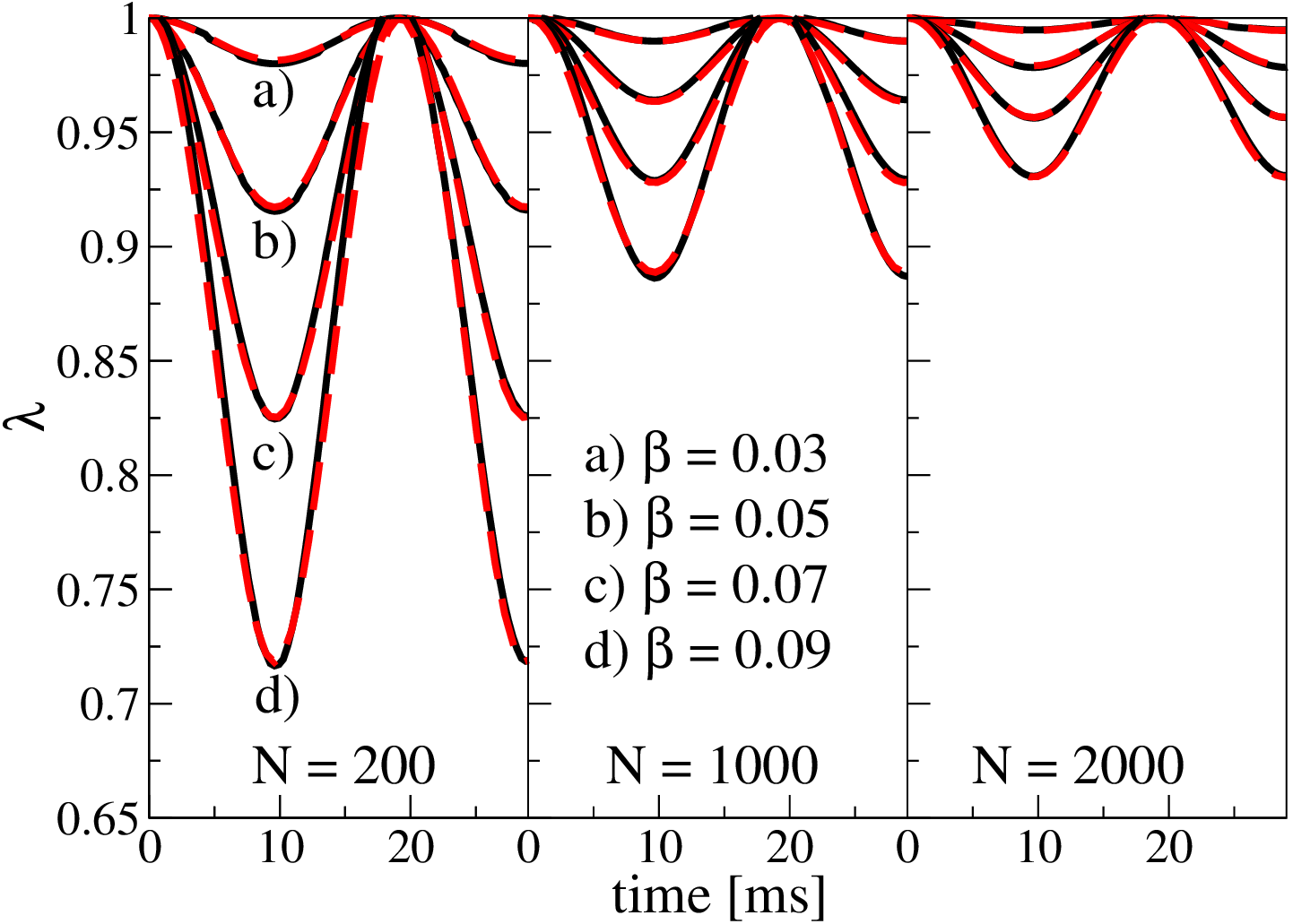}
\caption{Relative change of the BEC size after sudden switch-off of the dressing lasers ($\kappa$) for different BEC particle number and different dressing parameters. Red dashed lines correspond to self-similar density evolution ($\lambda$).}
\label{breathing}
\end{figure}
\begin{figure}
\centering
\includegraphics*[scale=0.30]{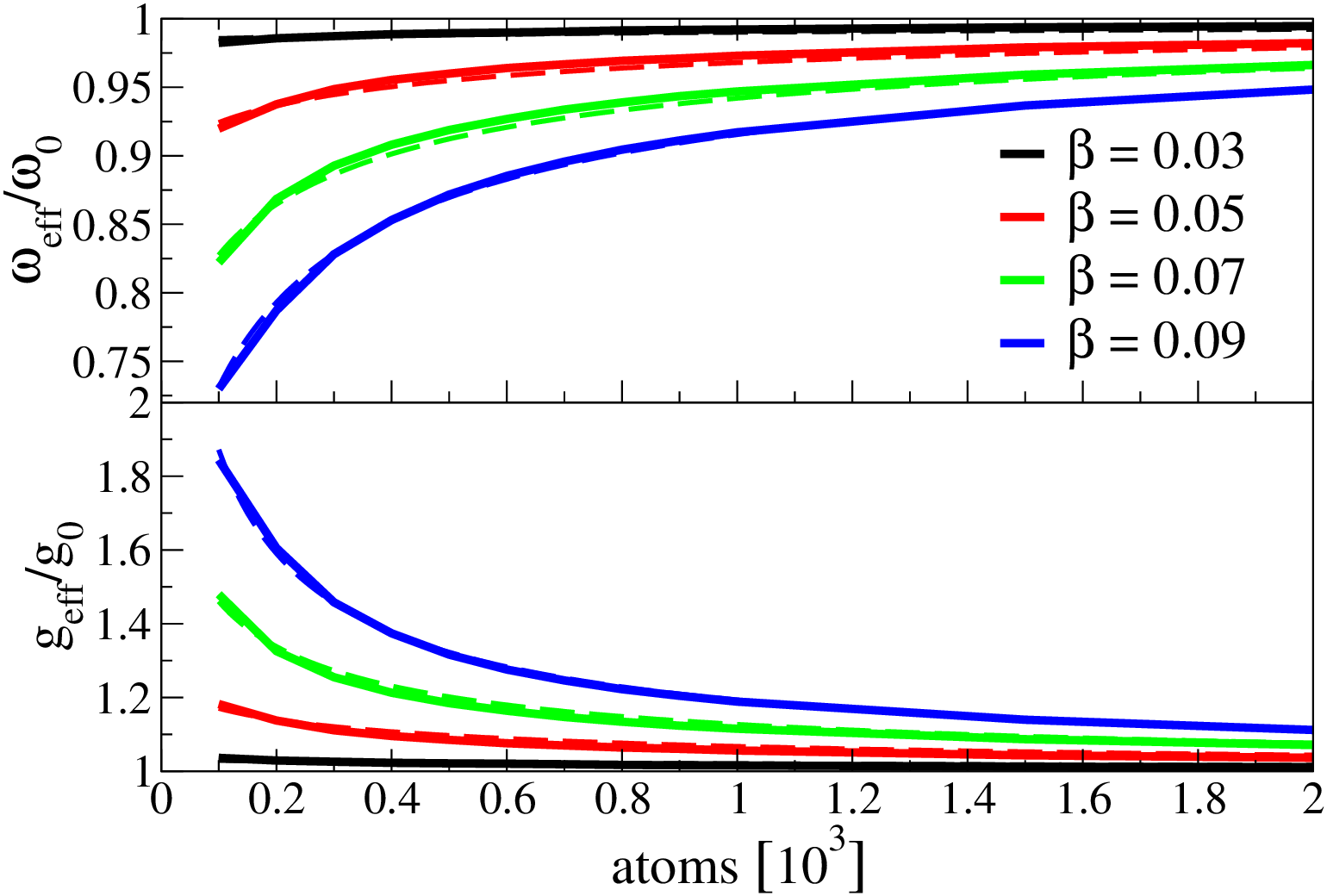}
\caption{Relative change of the effective trapping frequency $\omega_{\rm eff}$ (top panel) and relative change of the mean-field coupling $g_{\rm eff}$ (bottom panel) for different atom number $N$ for fixed dressing parameter (curves from top to bottom: $\beta = 0.03, 0.05, 0.07, 0.09$). Dashed lines correspond to Eq.~(\ref{omega_analytical}).
}
\label{g_eff}
\end{figure}

This non-adiabatic switch-off of the dressing results in the excitation of a BEC breathing mode. 
The dynamics of a BEC can be described by the collective motion of atoms with time-dependent density $\rho(x,t) \propto \rho_0\left(x/\lambda(t) \right)/\lambda(t)$, where $\rho_0(x)$ is the initial BEC density  and the scaling parameter $\lambda(t)$ obeys $\frac{d^2\lambda}{d t^2} = \frac{\omega_{\rm eff}^2}{\lambda^2} - \omega_0^2\lambda$ \cite{Dum1996}. A solution of this equation is periodic, with an amplitude  parameterized  by $\omega_{\rm eff}$, and 
a frequency $\omega_b\simeq \sqrt{3}\omega_0$. For fixed $N$, $n$, and $\beta$ we calculate $\kappa(t)$, which we then use to find $\omega_{\rm{eff}}$ by minimizing the root mean square of the difference between $\lambda(t)$ and $\kappa(t)$. Figure \ref{breathing} presents $\kappa$ (black solid lines) and $\lambda$ (red dashed lines) for fixed number of atoms $N = 200, 1000, 2000$ and fixed dressing parameter $\beta = 0.03, 0.05, 0.07, 0.09$, with corresponding critical densities $\rho_B$ equal to $177\,\mu$m$^{-1}$, $64\,\mu$m$^{-1}$, $33\,\mu$m$^{-1}$ and $20\,\mu$m$^{-1}$, respectively. Knowing $\omega_{\rm eff}$ we can extract the effective density-dependent change of the mean-field interaction strength $g_{\rm eff}/g_0 = (\omega_{\rm eff}/\omega_0)^{-2}$. Figure \ref{g_eff} presents the change of the effective trapping freqency and of the effective mean-field interaction strength as a function of particle number $N$, for fixed dressing parameter $\beta$, which agrees very well with the analytical expression of Eq.~(\ref{omega_analytical}).
 
An important consideration is the decay associated with the Rydberg dressing, including the effects of radiative decay of the intermediate state in the two-photon coupling scheme.
 In practice the atomic system is a three-level system with an atomic ground state $|g\rangle$ coupled to a Rydberg $S$-state $|r\rangle$ via an intermediate state $|e\rangle$. The Rabi frequency and detuning for the transition from $|g\rangle$ to $|e\rangle$ are $\varOmega_1$ and $\varDelta_1$, while similarly for the transition from the intermediate state $|e\rangle$ to the Rydberg state $|r\rangle$ they are denoted as $\varOmega_2$ and $\varDelta_2$. The intermediate level $|e\rangle= |5P_{3/2}\rangle$ is far detuned, i.e.: $|\varDelta_1|,\,|\varDelta_2|   \gg \varOmega_1, \varOmega_2, \varGamma_{5P_{3/2}}$ and can be adiabatically eliminated. This effectively reduces the three-level system to a two-level system 
with two-photon Rabi frequency $\varOmega =  \varOmega_{1}\varOmega_2/2|\varDelta_1|$ and total detuning $\varDelta = \varDelta_1 + \varDelta_2$, with $|\varDelta|\ll |\varDelta_1|,\,|\varDelta_2|$.

As an example, we can take the red (780-nm wavelength) laser parameters as having a waist of $r_{\rm red} = 500\mu$m with a power of $P_{\rm red} = 10\mu$W, and blue (480-nm wavelength) laser parameters as having a waist of $r_{\rm blue} = 90\mu$m with a maximum power of $P_{\rm blue} = 50$mW with fixed intermediate  state  detuning $\varDelta_1 = 2\pi\times 1.5$~GHz. Keeping the total detuning $\varDelta = -2\pi \times 100$~kHz constant as before, we change the dressing parameter $\beta$ via the Rabi frequency $\Omega_2$. 
We calculate the lifetime of  the  Rydberg dressed state by considering a weak admixture to the intermediate state $|5P_{3/2}\rangle$ with a decay rate $\varGamma_{5P_{3/2}}$, and an admixture of the Rydberg state
$|r\rangle = |nS\rangle$ with a decay rate $\varGamma_{|nS\rangle}$, which results in an effective decay rate  smaller than
$\varGamma_{\rm eff} = \beta^2\varGamma_{|nS\rangle} + ( \varOmega_1/2\varDelta_1)^2\varGamma_{5P_{3/2}}$.
(Note that the effective decay rate is smaller when collective effects play a role because in that case one should replace $\beta^2$ with $f$ and $f<\beta^2$).
The lifetime $\varGamma_{|nS\rangle}^{-1}$ for  the  $n=35S$ $^{87}\text{Rb}$ Rydberg state is $25.2\mu$s at a  temperature $T = 300$ K \cite{Beterov2009}. The intermediate state decay rate is $\varGamma_{5P_{3/2}}  = 2\pi \times 6.1$~MHz, and $\varOmega_1/2\varDelta_1= 18\times 10^{-4}$. The corresponding minimum life-times for  $\beta = 0.03, 0.05, 0.07, 0.09$ 
are then $6.3$ ms, $4.5$ ms, $3.1$ ms and $2.2$ ms respectively.
From these results, it is clear that it is realistic to observe Rydberg dressing in an experiment: for instance,  for the parameters $N=1000$, $\beta=0.07$, the resulting change in BEC size during the breathing is about 8\%. 

 It should be possible to reach the dressed ground state by turning on the dressing lasers during the very final stages of evaporative cooling. 
An alternative approach is to consider adiabatic turn-on  and rapid switch-off  of the dressing; this is beyond the scope of the present paper. 
An advantage of the above scheme is that the resulting breathing occurs in the non-dressed BEC, so that the limited lifetime of the dressed state is not an issue during the breathing.

The Rydberg-mediated control over the interactions we have discussed here offers an important alternative to previously considered schemes. For instance, Feshbach resonances allow for time-dependent non-linear dynamics as they can be utilized for a periodic modification of the mean field coupling. This was proposed by Saito and Ueda \cite{Ueda2003}, who considered a sinusoidal time-dependent modulation of the coupling constant, and by Kevrekidis \cite{Malomed2003}, who considered a block-type of periodic modulation of the mean-field coupling constant. An intriguing opportunity offered by the dressing is that by spatially modulating the dressing lasers (e.g.\ in a standing wave) it would be possible to create a {\em spatial} modulation of the interaction strength, as well as a temporal modulation, which can be rapidly switched. 

This research was financially supported by the Foundation for Fundamental Research on Matter (FOM), and by the Netherlands Organisation for Scientific Research (NWO). We also acknowledge the European Union H2020 FET Proactive project RySQ (grant N. 640378).


\end{document}